\documentclass{IEEEoj}
\usepackage{cite}
\usepackage{amsmath,amssymb,amsfonts}

\usepackage{graphicx,color}
\usepackage{textcomp}
\usepackage{graphicx}

\usepackage{graphicx}
\usepackage{amsmath}
\usepackage{amssymb}
\usepackage{booktabs}

\usepackage{graphicx}
\usepackage{algorithm}
\usepackage{algpseudocode}
\usepackage{float}
\usepackage{stfloats}
\usepackage{caption}
\usepackage{color} 
\usepackage[margin=0pt,position=bottom,font=footnotesize,labelfont=bf]{caption}
\usepackage[margin=0pt,position=bottom,font=footnotesize,labelfont=bf]{subcaption}
\usepackage[export]{adjustbox}
\usepackage{graphics,cite}
\usepackage{array}
\usepackage{amsmath,amsfonts}
\usepackage{siunitx}

\usepackage{flushend}
\usepackage{pict2e}
\usepackage{tikz}
\usepackage{wrapfig}

\usepackage[normalem]{ulem}

\newif\ifannotated
\annotatedtrue

\ifannotated

\newcommand{\delete}[1]{{\color{red}{\sout{#1}}}}

\newcommand{\margincomment}[1]{\marginpar{$\Rightarrow$\color{red}\fbox{\parbox{\linewidth}{\color{black}\scriptsize#1}}}}
\else

\newcommand{\delete}[1]{\ignorespaces}

\newcommand{\margincomment}[1]{}
\fi

\newcommand{\cm}{\text{cm}}

\def\BibTeX{{\rm B\kern-.05em{\sc i\kern-.025em b}\kern-.08em
    T\kern-.1667em\lower.7ex\hbox{E}\kern-.125emX}}
\AtBeginDocument{\definecolor{ojcolor}{cmyk}{0.93,0.59,0.15,0.02}}
\begin{document}
\receiveddate{XX Month, XXXX}
\reviseddate{XX Month, XXXX}
\accepteddate{XX Month, XXXX}
\publisheddate{XX Month, XXXX}
\currentdate{XX Month, XXXX}
\doiinfo{OJIM.2022.1234567}

\title{Coded Illumination for 3D Lensless Imaging}

\author{Yucheng Zheng and M. Salman Asif (Senior Member, IEEE)}
\affil{Department of Electrical and Computer Engineering, University of California, Riverside}
\corresp{CORRESPONDING AUTHOR: M. Salman Asif (e-mail: sasif@ucr.edu).}
\authornote{This work was supported in part by National Science Foundation (NSF) grants CCF-2046293 and CMMI-2133084.}
\markboth{Coded Illumination for 3D Lensless Imaging}{Zheng and Asif}

\begin{abstract}
 Mask-based lensless cameras offer a novel design for imaging systems by replacing the lens in a conventional camera with a layer of coded mask. Each pixel of the lensless camera encodes the information of the entire 3D scene. Existing methods for 3D reconstruction from lensless measurements suffer from poor spatial and depth resolution. This is partially due to the system ill conditioning that arises because the point-spread functions (PSFs) from different depth planes are very similar. In this paper, we propose to capture multiple measurements of the scene under a sequence of coded illumination patterns to improve the 3D image reconstruction quality. In addition, we put the illumination source at a distance away from the camera. With such baseline distance between the lensless camera and illumination source, the camera observes a slice of the 3D volume, and the PSF of each depth plane becomes more resolvable from each other. We present simulation results along with experimental results with a camera prototype to demonstrate the effectiveness of our approach. 
\end{abstract}



\maketitle

\section{Introduction} 
Lensless cameras provide novel designs for extreme imaging conditions that require small, thin form factor, large field-of-view, or large-area sensors \cite{asif2017flatcam,boominathan2016lensless,antipa2018diffusercam,boominathan2022recent}. Compared to conventional lens-based cameras, lensless cameras are flat, thin, light-weight, and geometry flexible.
Depth estimation with lensless imaging has been a challenging problem \cite{antipa2018diffusercam, adams2017rice_depth, zheng2019joint}. {The primary reason is that the sensor responses for different depth planes have small differences, which makes the 3D reconstruction an ill-conditioned problem.}

In this paper, we propose a new method that combines coded illumination with mask-based lensless cameras (such as FlatCam\cite{asif2017flatcam}) to improve the quality of recovered 3D scenes. We project a sequence of coded illumination patterns onto the 3D scene and capture multiple frames of lensless measurements. We then solve an inverse problem to recover the 3D scene volume using all the coded measurements. Coded illumination-based measurements provide {a better-conditioned system and improve the quality}  of 3D reconstruction. 
The illumination source is separated from the camera by a baseline distance, which ensures that the depth-dependent point spread functions (PSFs) of each depth plane is different from one another. 
The choice and design of the illumination source depend on the application of the imaging system. We use a projector installed next to the lensless camera as the illumination source. 

The main contributions of this paper are as follows. 
\begin{itemize}
\item We propose a novel framework to capture lensless measurements under a sequence of coded illumination patterns and improve the 3D reconstruction results.

\item We show that the baseline between projector and camera cause depth-dependent shifts of PSF and enhance the 3D performance at large distances.

\item We provide simulation and experimental results to validate the proposed method. Our experiments show that the quality of 3D reconstruction improves significantly with coded illumination. 

\end{itemize}

\section{Related Work}

Mask-based lensless cameras, such as FlatCam \cite{asif2017flatcam}, can be viewed as extended versions of pinhole cameras. Although a pinhole camera is able to image the scene directly on a sensor, it often suffers from severe sensor noise \cite{yedidia2018analysis_aperture}. Coded aperture-based cameras alleviate this problem by using multiple pinholes arranged in a designed pattern \cite{asif2017flatcam, fenimore1978ura, busboom1998ura, cannon1980coded_aperture, boominathan2016lensless}. The scene is reconstructed by solving an inverse problem using the linear multiplexed lensless measurements. With the small baseline between the pinholes on the mask, the coded aperture-based cameras are also able to capture the depth information of the scene \cite{levin2007_mask_on_lens,asif2018lensless, antipa2018diffusercam, adams2017rice_depth,zheng2019joint,hua2020sweep,zheng2021simple}. {3D reconstruction using a single snapshot of a lensless camera is an under-determined and highly ill-conditioned problem \cite{zheng2019joint}.}

Signal recovery from ill-conditioned and under-determined systems is a long-standing problem in signal processing.  
A standard approach to deal with ill-conditioned and under-determined systems is to add a signal-dependent regularization term in the recovery problem, which constrains the range of the solutions. Popular methods include adding sparse and low-rank priors \cite{candes2006compressive, rudin1992tv, donoho2006compressive, richard2007compressive,recht2010matrix} or natural image prior \cite{rakib2019generator, bora2017generator, hand2018phase}. 
Recently, a number of methods have been proposed that use deep networks to reconstruct or post-process the images from lensless measurements \cite{khan2020flatnet,boominathan2020phlatcam,kristina2019learning,kristina2021untrained}. Some of these methods provide exceptional improvement over traditional optimization-based methods. Nevertheless, deep learning-based methods in general, and end-to-end methods in particular, provide a huge variation in performance for simulated and real data (mainly because of mismatch in the simulated/actual mask-sensor-projector configuration and scenes). In contrast to deep learning methods, our method seeks to improve the conditioning of the underlying linear system and offer better generalization and robust results for arbitrary scenes without the need for learning from data \cite{hua2020sweep,zheng2021simple}.

Our proposed approach can be viewed as an active imaging approach combining coded modulation or structured illumination method with coded aperture imaging
\cite{gustafsson2008structured, nayar2012diffuse, fofi2004survey}.
Structured illumination schemes are commonly used for imaging beyond diffraction in microscopy. These schemes use multiple structured illumination patterns to down-modulate high spatial frequencies in a sample into a low-frequency region that can be captured by the microscope \cite{gustafsson2008structured,rainer1999SIM,gustafsson2000surpassing}. 
Coded illumination for lensless imaging of 2D scenes was recently presented in \cite{zheng2020coded-icassp,zheng2021coded}. 
Another active imaging approach uses time-of-flight sensors \cite{gokturk2004tof,heide2013tof} that estimate the 3D scene by sending out infrared light pulses and measuring the traveling time of their reflections.

\section{Methods}

\subsection{Imaging Model}

\begin{figure}[t]
    \centering
    \includegraphics[width=0.9\linewidth,keepaspectratio]{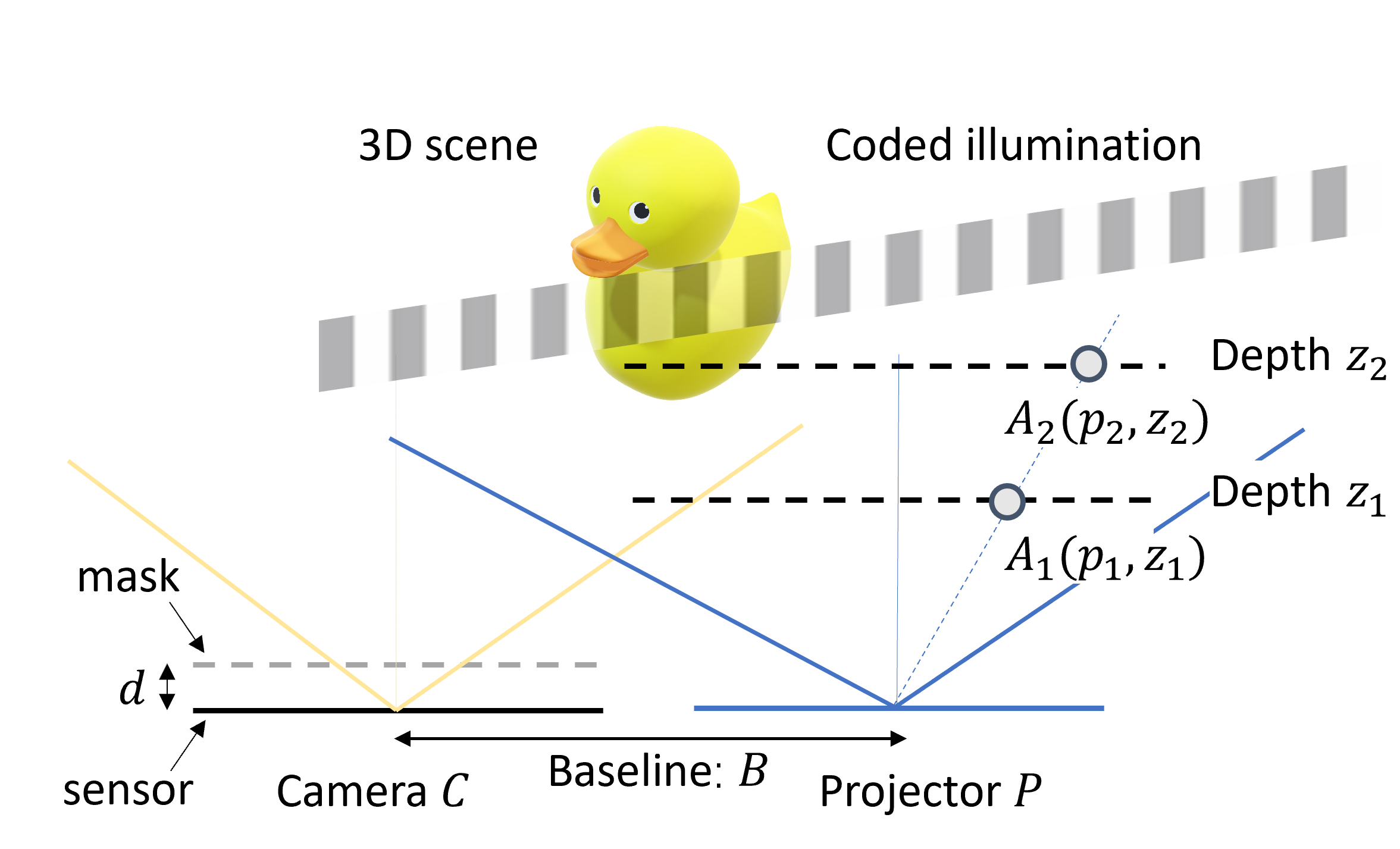}
    \caption{Illustration of a lensless camera with coded illumination. Camera and projector are separated by baseline distance $B$. 
    The 3D scene is illuminated by a sequence of coded illumination patterns from the projector, and observed by the camera sensor beneath the coded mask. 
    Rays that receive same illumination in projector coordinates appear at different angles in camera coordinates that provides different depth-dependent PSFs. }
    \label{fig:intro}
\end{figure}

Mask-based lensless cameras replace the lens with a layer of coded mask and capture linear multiplexed measurements with an image sensor. The mask pattern 
can be placed parallel to the sensor plane at distance $d$, as illustrated in Figure~\ref{fig:intro}.  
In general, we can model the  measurement recorded at a sensor pixel $(s_u,s_v)$ as a linear function of the scene intensity as 
\begin{equation}
y(s_u,s_v) = \int I(x,y, z) \, \varphi(s_u,s_v;x,y,z) \, dx dy dz, 
\label{eq:integration}
\end{equation}
where $I(x,y,z)$ denotes the image intensity at 3D location $(x,y,z)$ and $\varphi(s_u,s_v;x,y,z)$ denotes the point spread function (PSF) or the  sensor response recorded at $(s_u,s_v)$ in the sensor plane for a point source at $(x,y,z)$.

{The general system in \eqref{eq:integration} can be simplified depending on the system design and placement and pattern of the mask.} In our proposed method, we use a separable model proposed in \cite{asif2017flatcam}, where we use a rank-1 matrix as the amplitude mask. 
With the separable mask placed parallel to the image sensor, the PSF of an arbitrary point,  $\varphi(s_u,s_v;x,y,z)$, will be a rank-1 matrix, and the general model in \eqref{eq:integration} can be written in a simpler form as a  separable system.

Suppose we discretize the continuous scene $I(x,y,z)$ into $D$ depth planes $\mathbf{I_1,\ldots,I_D}$, each with $N\times N$ pixels. The separable system can be represented in the following compact form: 
\begin{equation}
\mathbf Y =  \sum_k \mathbf{\Phi_kI_k\Phi_k^T}.
\label{eq:separable_model}
\end{equation}
{$\mathbf{Y}$ represents $M\times M$ sensor measurements and $\Phi_k$ represents the system matrix for the $k$-th depth plane.}

\subsection{Coded Illumination}
{We use a projector separated by baseline distance $B$ from the camera to illuminate the scene with a sequence of coded illumination patterns (as illustrated in Figure~\ref{fig:intro}. }
The effect of coded illumination can be modelled as an element-wise product between the illumination pattern and the scene. We divide the field-of-view (FOV) cone of the projector into $N\times N$ angles, {which also determines the spatial discretization of the scene.} {We generate a sequence of illumination patterns and capture the corresponding measurements on the sensor.} 
The measurements captured for $i$-th pattern $\mathbf{P_i}$ can be represented as 
\begin{equation}
\mathbf Y_i =  \sum_k \mathbf{\Phi_k(P_i\odot I_k)\Phi_k^T}.
\label{eq:separable_model_illum}
\end{equation}
{Note that we assume the same illumination pattern for every depth plane at a time. This is because we use the projector to determine the scene discretization at every depth plane.}

To recover the 3D scene as a stack of $D$ planes,  {$\mathbf{I = \{I_1,\ldots, I_D\}}$}, we solve the following regularized least-squares problem: 
\begin{equation}
   \min_{\mathbf{I_1,\ldots, I_D}}~\sum_i\|\mathbf{Y_i}-\sum_k \mathbf{\Phi_k(P_i\odot I_k)\Phi_k^T}\|_2^2+\lambda \|\mathbf{D(I)}\|_2.
    \label{eq:least_squares}
\end{equation}
{$\mathbf{D(I)}$ represents a finite difference operator that computes local gradients of the 3D volume $\mathbf{I}$ along spatial and depth directions. The $\ell_2$ norm of the local differences provides the 3D total variation function that we use as the regularization function. The total variation function constrains the magnitude of the local variation in the reconstruction and is widely used in ill-conditioned image recovery problems~\cite{li2013tv, rudin1992tv}. The optimization problem in \eqref{eq:least_squares} can be solved using an iterative least-squares solvers; we used the TVreg package~\cite{jensen2012tvreg}. }

\subsection{Effect of Baseline on Depth-Dependent PSFs }

As discussed in previous work on 3D lensless imaging \cite{antipa2018diffusercam, adams2017rice_depth, hua2020sweep, zheng2019joint}, the points at different depth in the scene provide a scaled version of the mask pattern as the sensor response.  
However, if the object is far from the lensless camera, the depth-dependent differences in the sensor reponse become almost negligible. 

{Coded illumination in our proposed system provides robust 3D reconstruction for two main reasons: (1) Coded illumination selects a subset of scene points that contribute to each sensor measurement. (2) Spatial separation between camera and projector (i.e., baseline) maps depth variations in scene points into depth-dependent shifts in the sensor response. Since shifted versions of the the mask pattern can be easily resolved compare to the scaled versions, the baseline plays a critical role in quality of 3D reconstruction.} 

{Let us consider the 1D case of our proposed framework}, the projector {$P$ is placed at a baseline distance $B$ away from the camera $C$}, as shown in Figure~\ref{fig:intro}.
For an arbitrary point at $(p,z)$ in the coordinate system of $C$, its measurements on camera $C$ can be written as   
\begin{equation}
    \phi(s;p,z)=\text{mask}\left[(1-\frac{d}{z})s+d\frac{p}{z} \right],
        \label{eq:meas_pz}
\end{equation}
{where $s$ denotes the coordinates on the camera sensor and $d$ denotes the sensor-to-mask distance.  }
The coordinates of two arbitrary points ${A_1}(p_1,z_1)$ and ${A_2}(p_2,z_2)$ on the same ray of the projector in the coordinate system of $P$ become $(p_1+B,z_1)$, $(p_2+B,z_2)$ in the coordinate system of camera $C$ (because of the the baseline between camera and projector). 
{We can represent the camera response or PSF corresponding to each of these points as }
\begin{align}
    \phi(s;p_1,z_1)&=\text{mask}\left[(1-\frac{d}{z_1})s+\frac{dp_1}{z_1}-\frac{dB}{z_1} \right] \notag\\ &=\text{mask}\left[(1-\frac{d}{z_1})(s-\frac{dB}{z_1-d})+d\frac{p_1}{z_1} \right]  
    \label{eq:proj_point_meas1}
    \end{align}
    \begin{align}
    \phi(s;p_2,z_2)&=\text{mask}\left[(1-\frac{d}{z_2})s+\frac{dp_2}{z_2}-\frac{dB}{z_2} \right] \notag \\ &=\text{mask}\left[(1-\frac{d}{z_2})(s-\frac{dB}{z_2-d})+d\frac{p_2}{z_2} \right]. 
    \label{eq:proj_point_meas2}
\end{align}
Note that $\frac{p_1}{z_1}=\frac{p_2}{z_2}$ because the two points are at different depths on the same ray angle.
Therefore, the PSF of points $A_1$ and $A_2$ differ from each other with a scaling factor $1-\frac{d}{z}$ and a depth-dependent shift $\frac{dB}{z-d}$.

Specifically, when the object is far from the camera, we can often ignore the difference in depth scaling factor $1-\frac{d}{z}$, and the difference of the depth-dependent shift becomes $|\frac{dB}{z_1-d}-\frac{dB}{z_2-d}|.$
When the baseline is zero, two point light sources on the same ray are the scaled versions of each other, and the scaling factor becomes almost the same when the object distance $(z)$ is large. However, by separating the camera and project by baseline distance $B$, the camera observes a shifted 3D grid; two points on the shifted grid provide an angular difference with respect to the camera. Therefore, the depth resolvability of the system improves. This effect was previously discussed in \cite{asif2018lensless} for more general geometries with multiple cameras. 

{In general, the projector $P$ and camera $C$ can be separated laterally and axially. 
Lateral separation provides depth-dependent shifts of the PSF, which we discussed in \eqref{eq:proj_point_meas1} and \eqref{eq:proj_point_meas2}. Axial separation would provide depth-dependent scaling and shifts of the PSF, which can also be deduced from \eqref{eq:proj_point_meas1} and \eqref{eq:proj_point_meas2}. For instance, if the camera and projector are separated axially by $\Delta z$, the depth-dependent shifts can be calculated by replacing $z_1, z_2$  with $z_1+\Delta z, z_2+\Delta z$, respectively. Since these terms appear in the denominator, their influence on the PSF shift will be small compared to lateral baseline $B$.  }

\begin{figure}[t]
    \centering
        \begin{subfigure}[b]{1\linewidth}
    \centering
        	\setlength\tabcolsep{1pt}
	\renewcommand{\arraystretch}{1} 
	\begin{tabular}{ccccc}
        original image &
		uniform &
		16  lines &
		48  lines
		\\
		\includegraphics[width=0.24\linewidth,keepaspectratio]{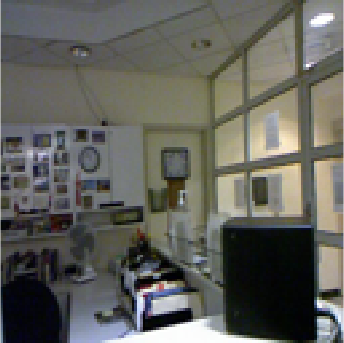} &
		\includegraphics[width=0.24\linewidth,keepaspectratio]{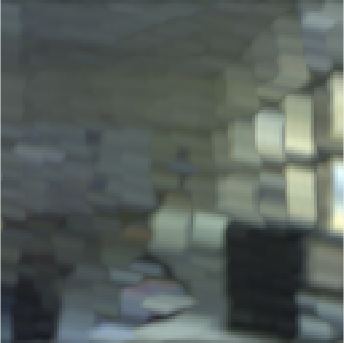} &
		\includegraphics[width=0.24\linewidth,keepaspectratio]{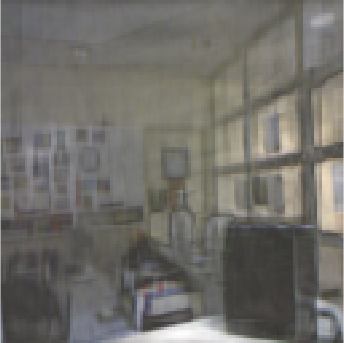} &
		\includegraphics[width=0.24\linewidth,keepaspectratio]{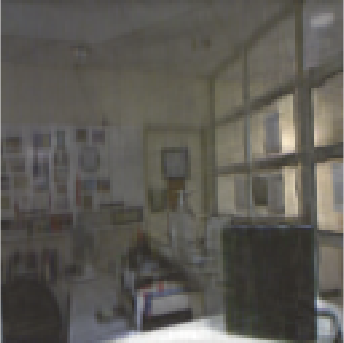} &
		\\
     original depth &
		SSIM: 0.28 &
		0.64 & 
		0.66
		\\
		\includegraphics[width=0.24\linewidth,keepaspectratio]{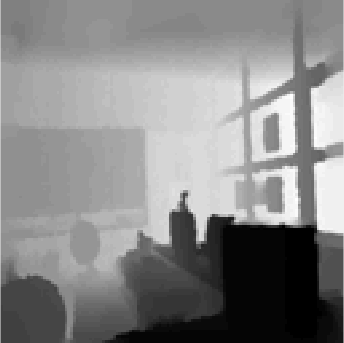} &
		\includegraphics[width=0.24\linewidth,keepaspectratio]{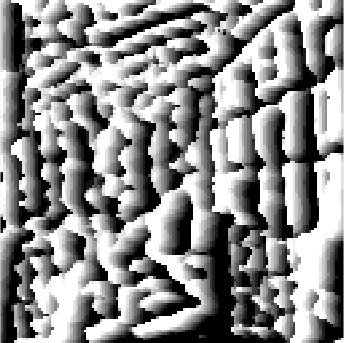} &
		\includegraphics[width=0.24\linewidth,keepaspectratio]{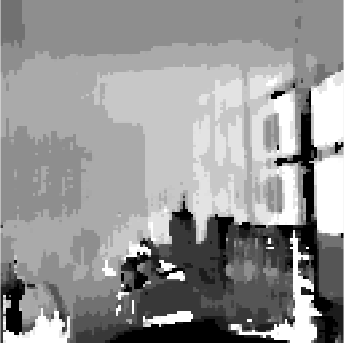}  &
		\includegraphics[width=0.24\linewidth,keepaspectratio]{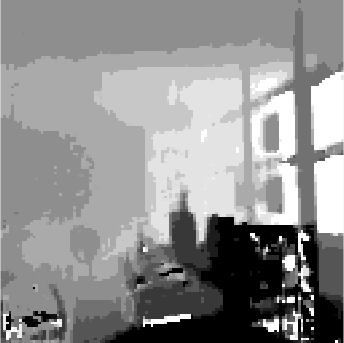} &
		\\
  &
		RMSE:8.28cm &
		4.53cm &
		2.96cm
	\end{tabular}
    \caption{Reconstruction of synthetic 3D test scene {for different numbers of illumination patterns with the same baseline. Top row represents the estimated all-in-focus images. Bottom row represents the estimated depth maps.}  }
	\end{subfigure}
	\begin{subfigure}[b]{0.9\linewidth}
	    \includegraphics[width=1\linewidth,keepaspectratio]{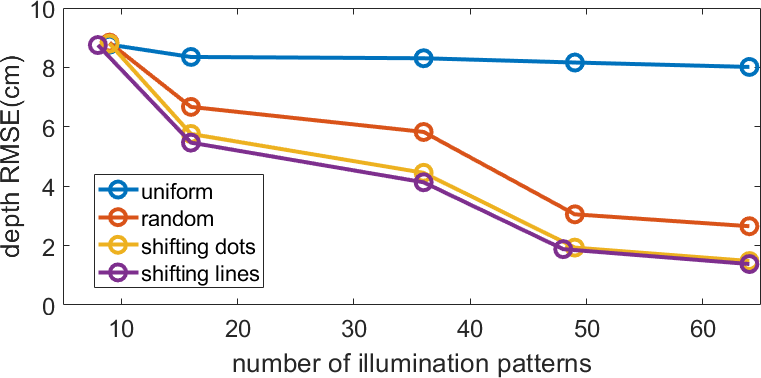}
	     \caption{Averaged depth RMSE of all test scenes. }
	\end{subfigure}
    \caption{Reconstruction and averaged depth RMSE for different number and types of illumination patterns. The baseline is fixed at 5cm during simulation. We observe that performance improves as we increase the number of illumination patterns. }
    \label{fig:simu_illumPattern}
\end{figure}

\begin{figure}[t]
    	\setlength\tabcolsep{1pt}
	\renewcommand{\arraystretch}{1} 
        \begin{subfigure}[b]{1\linewidth}
    \centering
	\begin{tabular}{ccccc}
        original image &
		{Baseline:} 0cm &
		1cm &
		5cm
		\\
		\includegraphics[width=0.24\linewidth,keepaspectratio]{figures/img_gt.png} &
		\includegraphics[width=0.24\linewidth,keepaspectratio]{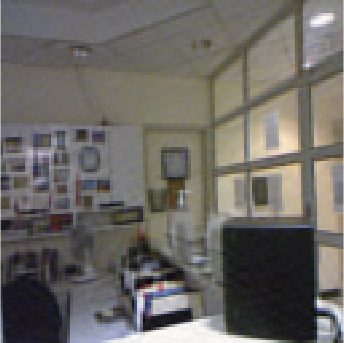} &
		\includegraphics[width=0.24\linewidth,keepaspectratio]{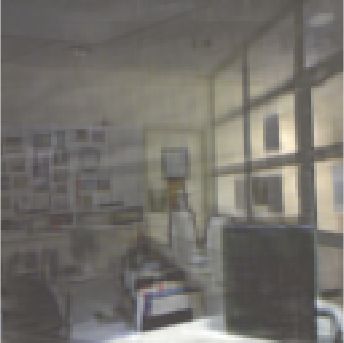} &
		\includegraphics[width=0.24\linewidth,keepaspectratio]{figures/img_baseline05_illum04_48.png} &
		\\
        original depth &
		SSIM: 0.91 &
		0.63 & 
		0.66
		\\
		\includegraphics[width=0.24\linewidth,keepaspectratio]{figures/depth_gt.png} &
		\includegraphics[width=0.24\linewidth,keepaspectratio]{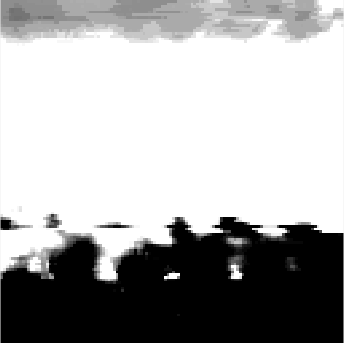} &
		\includegraphics[width=0.24\linewidth,keepaspectratio]{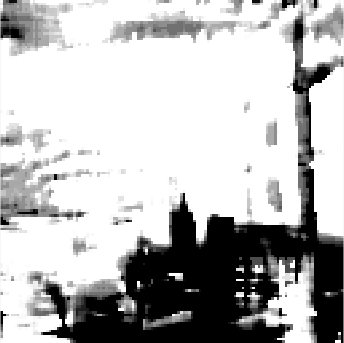}  &
		\includegraphics[width=0.24\linewidth,keepaspectratio]{figures/depth_baseline05_illum04_48.png} &
		\\
        &
		RMSE:8.57cm &
		6.60cm &
		2.96cm
	\end{tabular}
    \caption{Reconstruction of synthetic 3D test scene {for different baselines with 48 shifting lines patterns. Top row represents the estimated all-in-focus images. Bottom row represents the estimated depth maps.}  }
	\end{subfigure}
	\begin{subfigure}[b]{0.9\linewidth}
	    \centering
	    \includegraphics[width=1\linewidth,keepaspectratio]{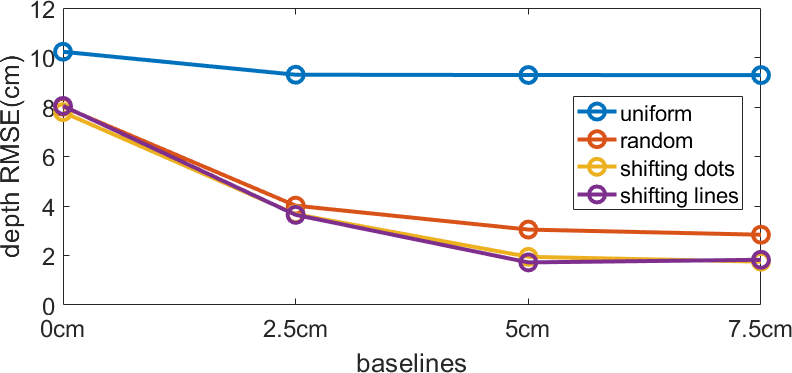}
	     \caption{Averaged depth RMSE of all test scenes. }
	     	    \label{fig:simu_baselines_plots}
	\end{subfigure}
    \caption{Reconstruction and averaged depth RMSE for different values of baseline distance. The number of illumination patterns is fixed for all tests. We observe that larger baselines provide better 3D reconstruction.}
    \label{fig:simu_baselines}
\end{figure}

\section{Simulation Results}

To validate the performance of the proposed algorithm, we simulate a lensless imaging system where a coded-mask is placed on top of an image sensor. We use a separable maximum length sequence (MLS) mask pattern \cite{busboom1998ura,asif2017flatcam}. The size of each mask feature is 60$\mu$m, and the sensor-mask distance is 2mm. The sensor pitch in the simulation is 4.8$\mu$m and the total number of pixels on the sensors is $512\times512$. We simulate a multi-plane 3D scene with $128\times128\times10$ voxels.
The simulated sensor noise consists of photon noise and read noise, and the noisy sensor measurements can be described as 
\begin{equation}
    \mathbf{Y}_n = \frac{G}{F}(\text{Poisson}(\frac{F}{G}\mathbf{Y})+N(0,\sigma^2)),
\end{equation}
where $\mathbf{Y}$ and $\mathbf{Y}_n$ refers to original and noisy measurements, where $F$ stands for the full-well capacity of the sensor, and $G$ represents the gain value. The variance $\sigma=F\times10^{-R/20}$ and $R$ is the dynamic range.

\subsection{Effect of Illumination Patterns}

%
We test different types of binary illumination patterns for the simulation. The patterns are designed to be binary to keep the model simple and to avoid the effect of {non-linearity caused by the Gamma curve of the projector.}. \\
\textbf{Uniform.} One pattern that illuminates all the pixels simultaneously; \\
\textbf{Random.} A sequence of separable binary random matrices. We ensure that the union of all the patterns should illuminate all the pixels (i.e., if we add up all the illumination patterns, they should not have zero entries anywhere). \\
\textbf{Shifting dots array.} The base illumination pattern consists of dots separated by $k$ pixels along the horizontal and vertical directions. We then generate a total of $k^2$ illumination patterns, each of which is a shifted version of the base pattern. The summation of all the patterns will give us a uniform illumination pattern. \\
\textbf{Shifting lines.} Similar to shifting dots array, the base illumination patterns consist of horizontal lines separated by $k$ pixels along vertical axis and vertical lines separated along horizontal axis. We then generate shifted version of these two base patterns. The summation of all the patterns is a uniform illumination pattern.

We present simulation results with different number and types of illumination patterns in Figure~\ref{fig:simu_illumPattern}. The simulated test scene is taken from NYU depth dataset\cite{silberman2012NYU_depth}. The depth of scene is rescaled into the range from $40\cm$ to $60\cm$ and discretized into 50 depth planes to simulate the sensor measurements. The camera setup and the baseline between camera and projector are fixed during the simulation. The shifting lines and shifting dots outperform the uniform pattern in terms of depth RMSE. Also, the depth RMSE drops as we increase the number of illumination patterns.

\newcommand{\figwidth}{0.24\linewidth}
\begin{figure}[t]
        \begin{subfigure}[b]{1\linewidth}
        \setlength\tabcolsep{1pt}
	\renewcommand{\arraystretch}{1} 
	\begin{tabular}{ccccc}
		MLS mask & pinhole & MLS mask &  pinhole \\
        16  lines & 16  lines & 48  lines & 48  lines
		\\
		\includegraphics[width=\figwidth,keepaspectratio]{figures/img_baseline05_illum04_16.png} &
		\includegraphics[width=\figwidth,keepaspectratio]{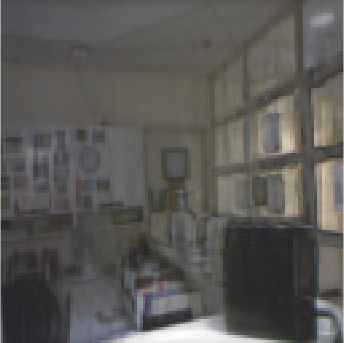} &
		\includegraphics[width=\figwidth,keepaspectratio]{figures/img_baseline05_illum04_48.png} &
		\includegraphics[width=\figwidth,keepaspectratio]{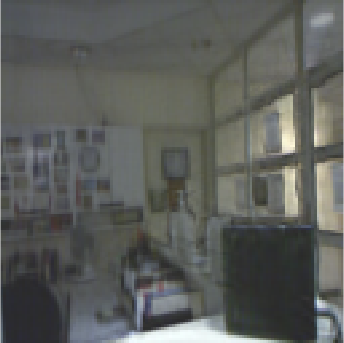} 
		\\
		SSIM: 0.64 & 
		0.66 &
		0.66 &
		0.69
		\\
		\includegraphics[width=\figwidth,keepaspectratio]{figures/depth_baseline05_illum04_16.png}  &
		\includegraphics[width=\figwidth,keepaspectratio]{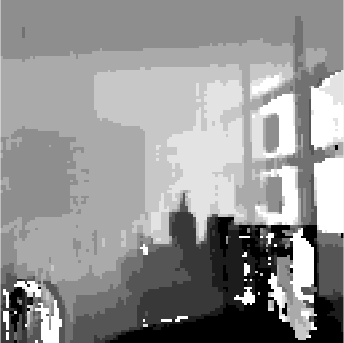} &
		\includegraphics[width=\figwidth,keepaspectratio]{figures/depth_baseline05_illum04_48.png} &
		\includegraphics[width=\figwidth,keepaspectratio]{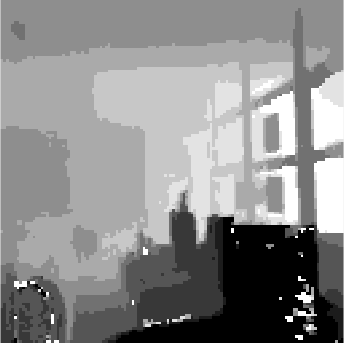} 
		\\
		RMSE: 4.53cm &
		3.52cm &
		2.96cm &
		1.72cm
	\end{tabular}
    \caption{Reconstruction of synthetic 3D test scene. {Top row represents the estimated all-in-focus images. Bottom row represents the estimated depth maps.}}
	\end{subfigure}
	\begin{subfigure}[b]{1\linewidth}
	    \centering
	    \includegraphics[width=1\linewidth,keepaspectratio]{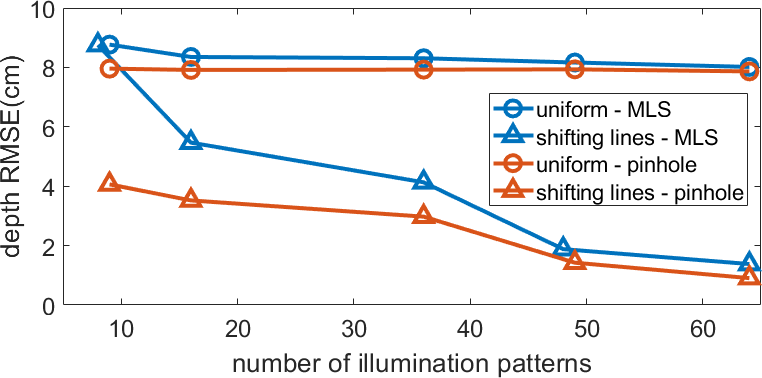}
	     \caption{Averaged depth RMSE of all test scenes.  }
	\end{subfigure}
    \caption{Comparison of the ideal pinhole-based and MLS mask-based camera models with coded illumination patterns. {The pinhole-based model performs better due to its better system conditioning. }  }
    \label{fig:simu_single_pinhole}
\end{figure}

\renewcommand{\figwidth}{0.23\linewidth}
\begin{figure}[t]
    \centering
        \begin{subfigure}[b]{1\linewidth}
    \centering
        	\setlength\tabcolsep{1pt}
	\renewcommand{\arraystretch}{1} 
	\begin{tabular}{ccccc}
		16 coded &
		16 mask &
		48 coded &
		48 mask \\
        illuminations & shifts \cite{hua2020sweep}  & illuminations & shifts \cite{hua2020sweep} \\
		\includegraphics[width=\figwidth,keepaspectratio]{figures/img_baseline05_illum04_16.png} &
		\includegraphics[width=\figwidth,keepaspectratio]{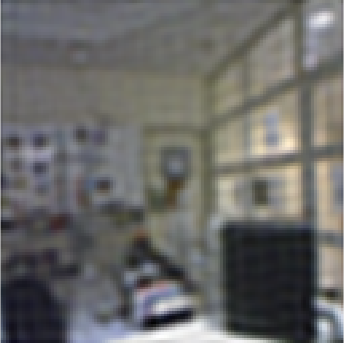} &
		\includegraphics[width=\figwidth,keepaspectratio]{figures/img_baseline05_illum04_48.png} &
		\includegraphics[width=\figwidth,keepaspectratio]{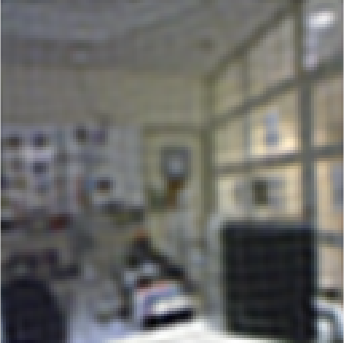} 
		\\
		SSIM: 0.64 & 
		0.61 &
		0.66 &
		0.62
		\\
		\includegraphics[width=\figwidth,keepaspectratio]{figures/depth_baseline05_illum04_16.png}  &
		\includegraphics[width=\figwidth,keepaspectratio]{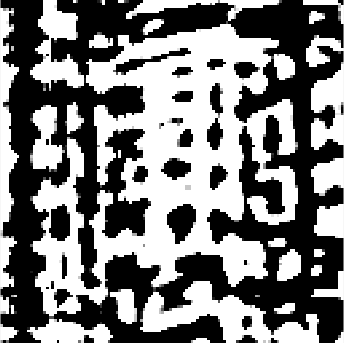} &
		\includegraphics[width=\figwidth,keepaspectratio]{figures/depth_baseline05_illum04_48.png} &
		\includegraphics[width=\figwidth,keepaspectratio]{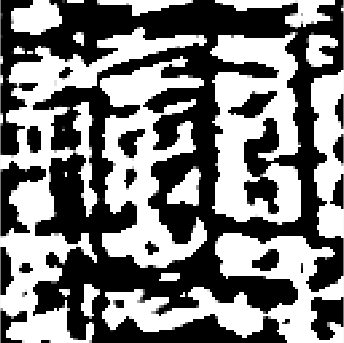} 
		\\
		RMSE: 4.53cm &
		10.48cm &
		2.96cm &
		9.87cm
	\end{tabular}
	\end{subfigure}
    \caption{Comparison of the proposed coded illumination-based reconstruction with shifting mask-based reconstruction in SweepCam~\cite{hua2020sweep}. The SweepCam fails to resolve objects that are far from the camera.  }
    \label{fig:simu_sweepcam}
\end{figure}

\subsection{Effects of Baseline}
The baseline between the lensless camera and the projector affects the depth resolvability of the system. Shifting the lensless camera by a distance, the camera observes the scene from a side view and transfers the depth difference of two points into angular difference. We present simulation results in Figure~\ref{fig:simu_baselines} to demonstrate the effect of camera-projector baselines. The number of illumination patterns for all the simulation are the same. We then fix the baseline along axial direction to 0cm and the baseline along lateral direction as $B=\{0, 2.5, 5, 7.5\}\cm$.  As shown in Figure~\ref{fig:simu_baselines}, the depth RMSE is decreased as we increase the baseline between the camera and the projector. When the baseline is zero, which means the camera and projector are overlapped, we barely distinguish any depth. 

{One important consideration for our method is that the target object should lie within the intersection of the sensor FOV cone and the projector illumination cone. As we increase the baseline, the
intersection of the two cones is pushed farther from the sensor. Therefore, we should determine the maximum baseline based on the object distance, sensor FOV, and projector cone. If we increase the baseline beyond the maximum limit, then the reconstruction quality can decrease. }

\subsection{Comparison with an Ideal Pinhole Camera}
In existing structured illumination methods \cite{gustafsson2008structured,rainer1999SIM,gustafsson2000surpassing}, a lens-based camera is used to capture the scene from the side view of the projector and depth map can be accurately reconstructed by triangulation. We can model the lens-based camera as an ideal pinhole camera (ignoring photon or sensor noise) for the sake of comparison with our method. We present simulation results comparing a pinhole-based camera with structured illumination in Figure~\ref{fig:simu_single_pinhole}. The baseline between the projector and the camera is fixed at 5cm in all the simulations. Results in Figure~\ref{fig:simu_single_pinhole} show that the pinhole mask (that represents an ideal lens-based camera) provides better results compared to the MLS mask. Compared to mask-based lensless camera where the sensor measurements are multiplexed, a lens-based system can offer better conditioning and depth reconstruction. Nevertheless, a lens-based camera imposes additional burden in terms of device thickness, weight, and geometry.

\begin{figure}[t]
    \centering
    \includegraphics[width=1\linewidth,keepaspectratio]{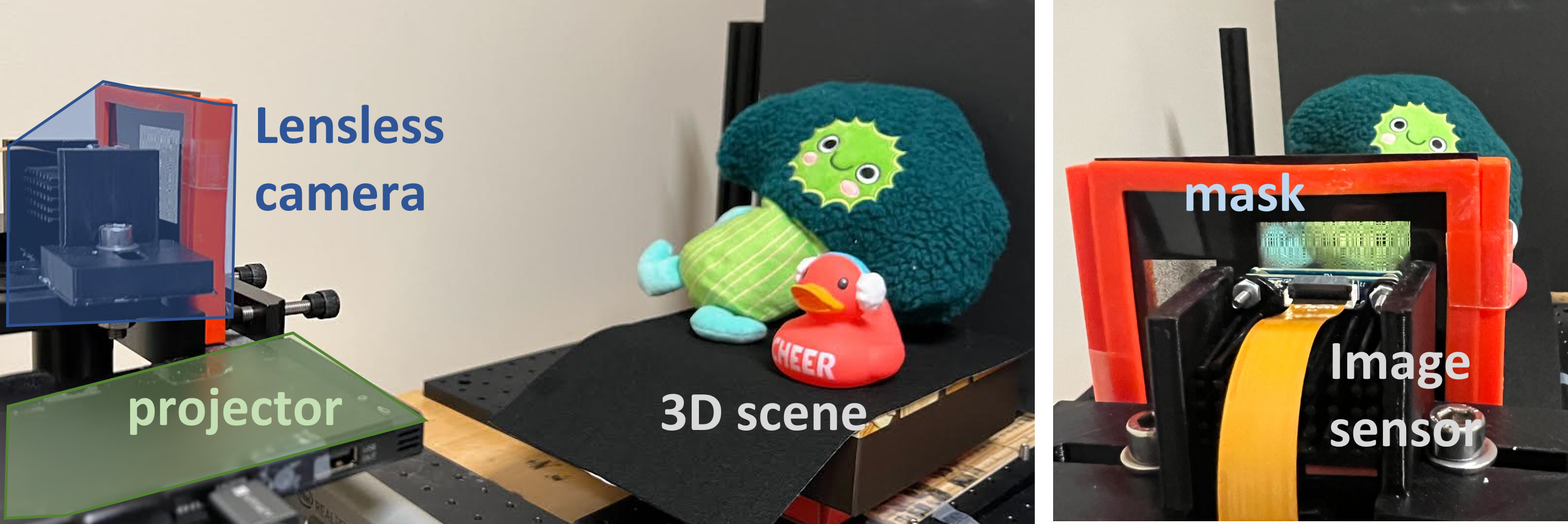}
    \caption{{Camera and projector setup used in our experiments. The projector is placed next to the camera. The scene objects are placed ranging from 40cm to 60cm. We capture multiple frames of sensor measurements under a sequence of coded illumination patterns from the projector to improve the 3D image reconstruction quality. } 
    }
    \label{fig:hardware_setup}
\end{figure}

\subsection{Comparison with Multishot Lensless System}

In our proposed method, multiple frames of measurements are captured, which introduce additional limitations such as long capture time and low frame rate.
In Figure~\ref{fig:simu_sweepcam}, we present simulation results comparing our method with another multi-shot lensless imaging system called SweepCam~\cite{hua2020sweep}. {SweepCam captures multiple frames of sensor measurements while translating the mask laterally. The translation of the mask offers a perspective shift in the measurements that depends on the depth of objects in the scene. }
In our simulations, the SweepCam mask is translated to 48 positions within a range of $2.88\text{mm}\times 2.88 \text{mm}$. However, since the translating distance of the mask is limited by the sensor area, the SweepCam method fails to resolve the depths when the scene is farther than 30cm.
 
\renewcommand{\figwidth}{0.13\linewidth}
\begin{figure*}[t]
    \centering
        	\begin{tabular}{ccccc|c}
		58cm &
		62cm &
		66cm &
		all-in-focus image &
		reconstructed depth &
        original~
		\\
		\rotatebox{90}{\parbox{2.4cm}{\centering  uniform}}
		\includegraphics[width=\figwidth,keepaspectratio]{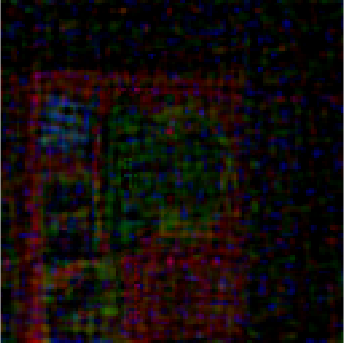} &
		\includegraphics[width=\figwidth,keepaspectratio]{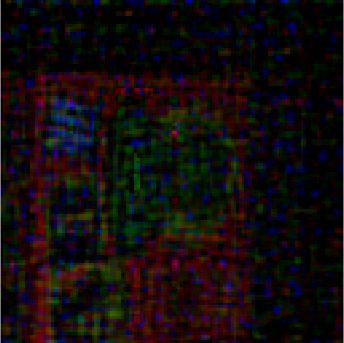} &
		\includegraphics[width=\figwidth,keepaspectratio]{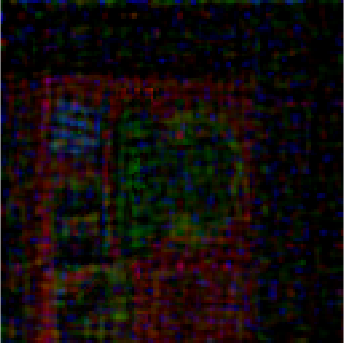} &
		\includegraphics[width=\figwidth,keepaspectratio]{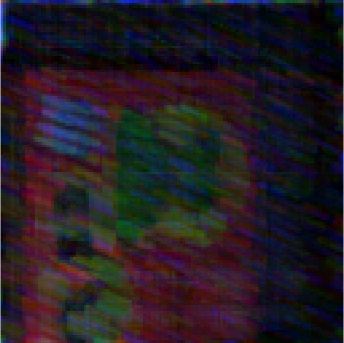} &
		\includegraphics[width=0.15\linewidth,keepaspectratio]{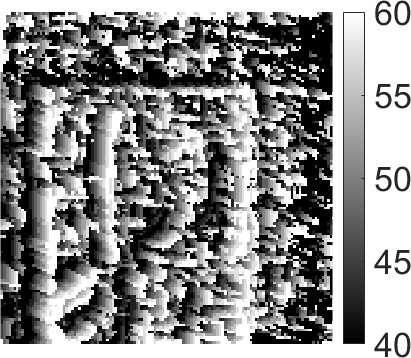} &
  		\includegraphics[width=\figwidth,keepaspectratio]{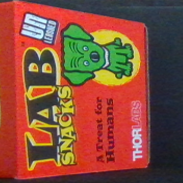} 
		\rotatebox{90}{\parbox{2.1cm}{\centering  image}}
		\\
		\rotatebox{90}{\parbox{2.4cm}{\centering 48 shifting lines}}
		\includegraphics[width=\figwidth,keepaspectratio]{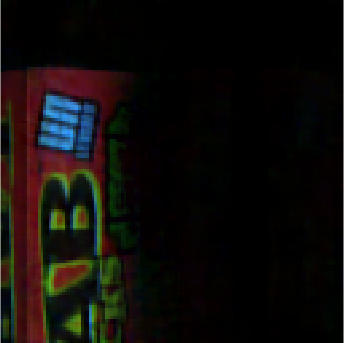}  &
		\includegraphics[width=\figwidth,keepaspectratio]{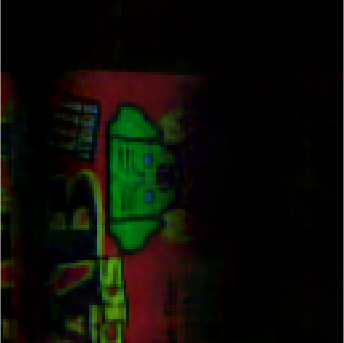} &
		\includegraphics[width=\figwidth,keepaspectratio]{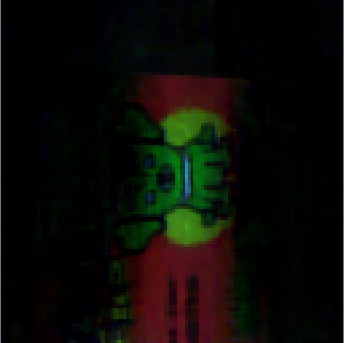} &
		\includegraphics[width=\figwidth,keepaspectratio]{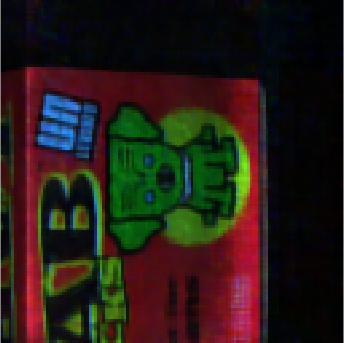} &
		\includegraphics[width=0.15\linewidth,keepaspectratio]{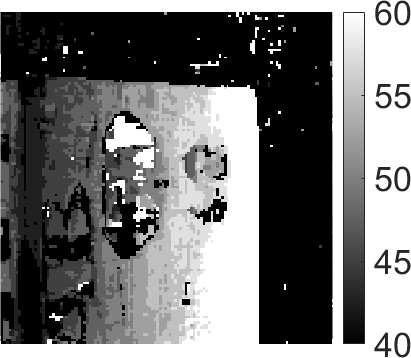} &
		\includegraphics[width=0.15\linewidth,keepaspectratio]{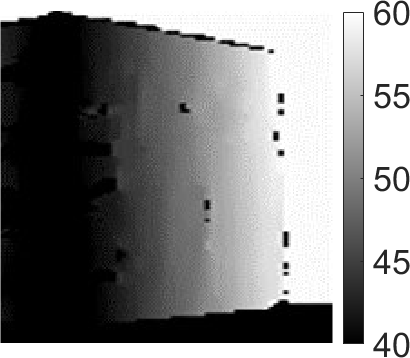}
  		\rotatebox{90}{\parbox{2.1cm}{\centering  depth}}
		\\
		\rotatebox{90}{\parbox{2.4cm}{\centering  uniform}}
		\includegraphics[width=\figwidth,keepaspectratio]{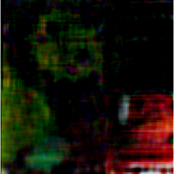} &
		\includegraphics[width=\figwidth,keepaspectratio]{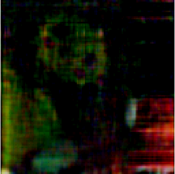} &
		\includegraphics[width=\figwidth,keepaspectratio]{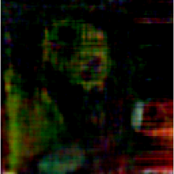} &
		\includegraphics[width=\figwidth,keepaspectratio]{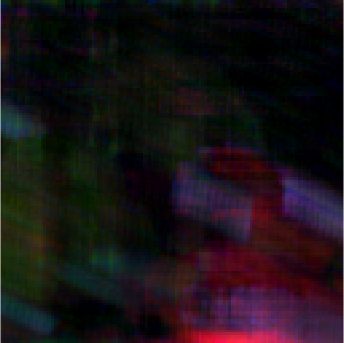} &
		\includegraphics[width=0.15\linewidth,keepaspectratio]{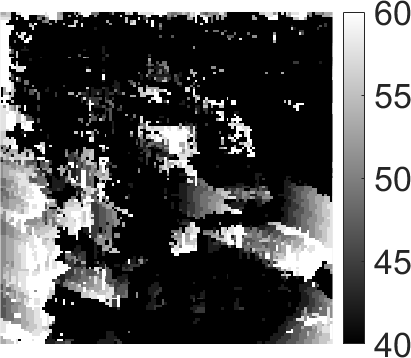} &
		\includegraphics[width=\figwidth,keepaspectratio]{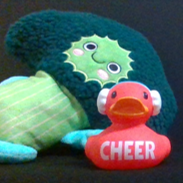} 
  		\rotatebox{90}{\parbox{2.1cm}{\centering  image}}
		\\
		\rotatebox{90}{\parbox{2.4cm}{\centering 48 shifting lines}}
		\includegraphics[width=\figwidth,keepaspectratio]{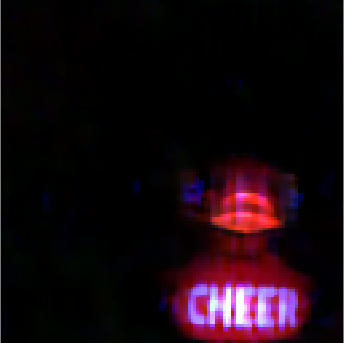}  &
		\includegraphics[width=\figwidth,keepaspectratio]{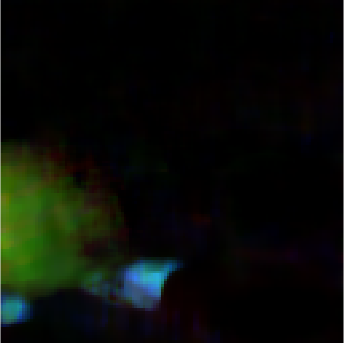} &
		\includegraphics[width=\figwidth,keepaspectratio]{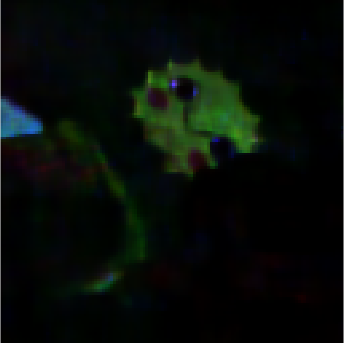} &
		\includegraphics[width=\figwidth,keepaspectratio]{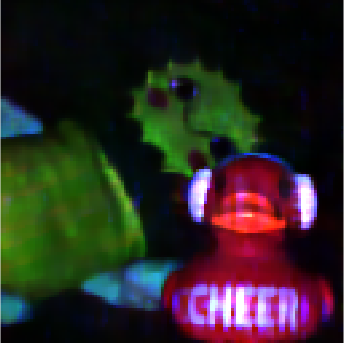} &
		\includegraphics[width=0.15\linewidth,keepaspectratio]{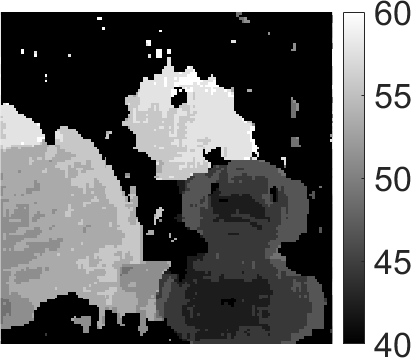} &
		\includegraphics[width=0.15\linewidth,keepaspectratio]{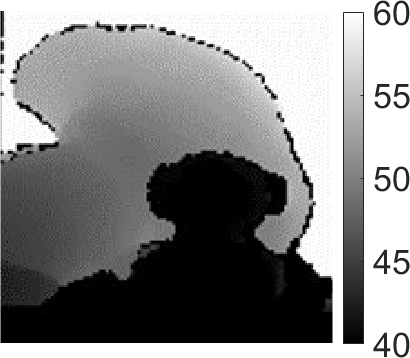} 
  		\rotatebox{90}{\parbox{2.1cm}{\centering  depth}}
	\end{tabular}
    \caption{Reconstructed images at three selected depth planes (58cm, 62cm, and 66cm), all-in-focus images, and depth maps using uniform and 48 shifting lines patterns. The depth maps of the real scenes and the estimated depth maps are all plotted in grayscale to show range from 40cm to 60cm. We observe that uniform illumination-based system fails to recover correct depth planes whereas the coded illumination-based system can recover  depth planes and entire 3D image with high quality. }
    \label{fig:exp_depthplanes}
\end{figure*}

\section{Experimental Results}
To validate our proposed method, we built a prototype with a lensless camera and a Sony MP-CL1 laser projector, {shown in Figure~\ref{fig:hardware_setup}}. The lensless camera prototype consists of an image sensor and a coded amplitude mask on top of it. We employ the outer-product of two MLS vectors as our mask pattern. The mask has $511\times 511$ square features. The pixel pitch is 60$\mu$m and the sensor-to-mask distance is 2mm. We use a Sony IMX183 sensor and bin $2\times 2$ sensor pixels, which yields the effective sensor pitch close to 4.8$\mu$m. We record $512\times512$ measurements from the sensor and the effective sensor size is $2.46\text{mm}\times2.46\text{mm}$. We place the test 3D objects within 40cm and 60cm depth range with respect to the camera.
Finally, we reconstruct $128\times128\times10$ voxels in the illuminated area. In our method, the lensless camera and the projector are separated by a 55mm baseline. We first reconstruct the depth planes by solving the regularized least-squares problem in \eqref{eq:least_squares}. Then we create an all-in-focus image and depth map by selecting the pixel with the maximum amplitude along each light ray.

{In our experiments, the pixel grid of the scene, illumination patterns $P_i$ , the system matrices at each depth $\mathbf{\Phi_k}$ must be correctly aligned; otherwise, we will get artifacts in the reconstruction. To avoid any grid mismatch, we use the same projector to calibrate the system matrices and generate the illumination patterns in our experiments.  }

\renewcommand{\figwidth}{0.12\linewidth}
\begin{figure*}[t]
    \centering
        	\begin{tabular}{ccccccc}
    	& original &
		uniform &
		16 shifting dots &
		49 shifting dots &
		16 shifting lines &
		48 shifting lines
		\\
		\rotatebox{90}{\parbox{2.2cm}{\centering  image}} &
		\includegraphics[width=\figwidth,keepaspectratio]{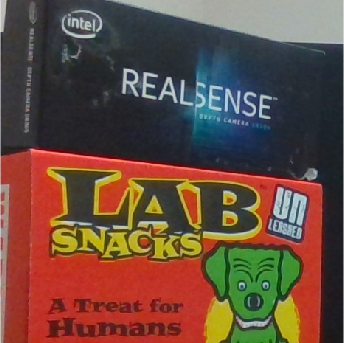} &
		\includegraphics[width=\figwidth,keepaspectratio]{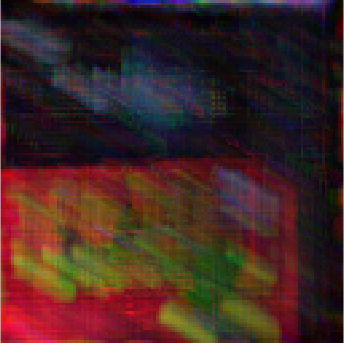} &
		\includegraphics[width=\figwidth,keepaspectratio]{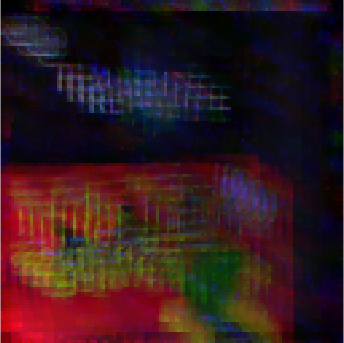} &
		\includegraphics[width=\figwidth,keepaspectratio]{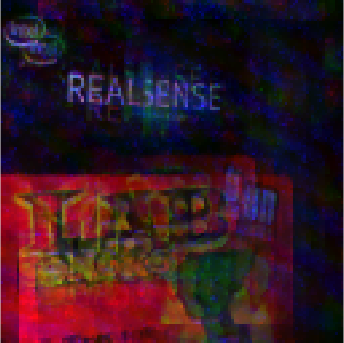} &
		\includegraphics[width=\figwidth,keepaspectratio]{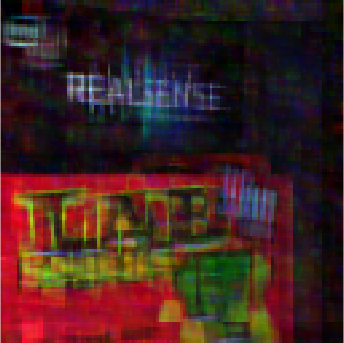} &
		\includegraphics[width=\figwidth,keepaspectratio]{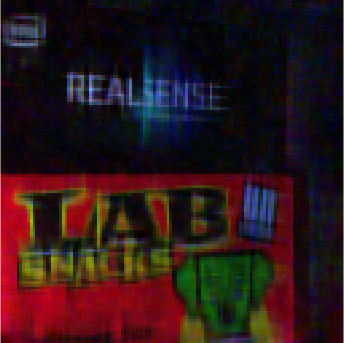} 
		\\
		\rotatebox{90}{\parbox{2.2cm}{\centering  depth}} & 
		\includegraphics[width=\figwidth,keepaspectratio]{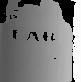} &
		\includegraphics[width=\figwidth,keepaspectratio]{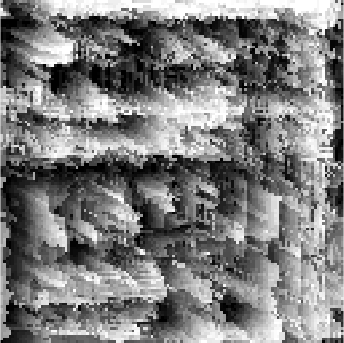} &
		\includegraphics[width=\figwidth,keepaspectratio]{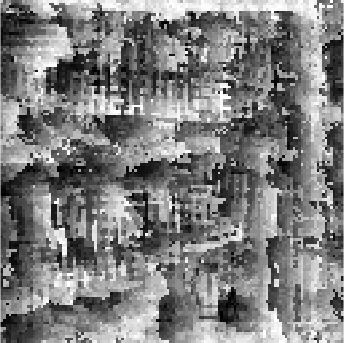}  &
		\includegraphics[width=\figwidth,keepaspectratio]{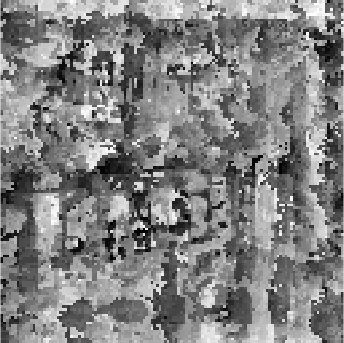}  &
		\includegraphics[width=\figwidth,keepaspectratio]{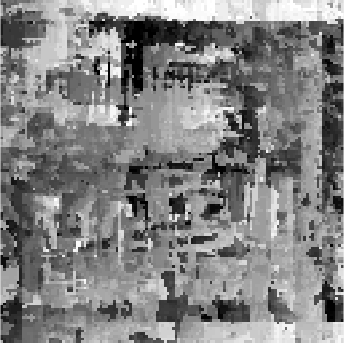} &
		\includegraphics[width=\figwidth,keepaspectratio] {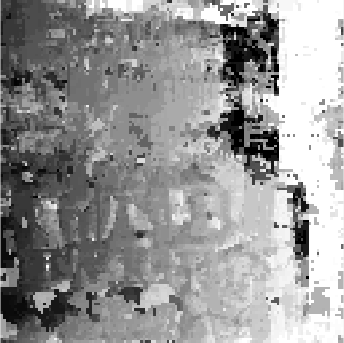}  
		\\
		\rotatebox{90}{\parbox{2.2cm}{\centering image}}
		& \includegraphics[width=\figwidth,keepaspectratio]{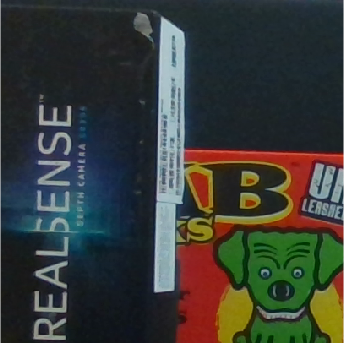} &
		\includegraphics[width=\figwidth,keepaspectratio]{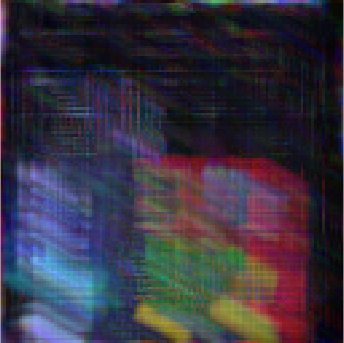} &
		\includegraphics[width=\figwidth,keepaspectratio]{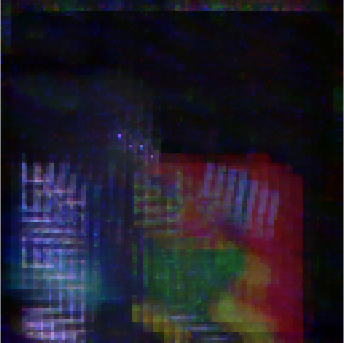} &
		\includegraphics[width=\figwidth,keepaspectratio]{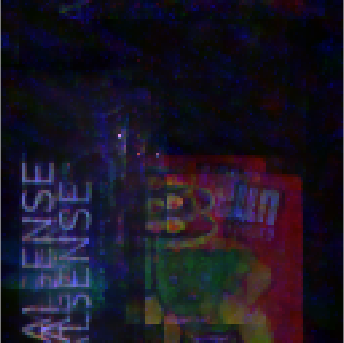} &
		\includegraphics[width=\figwidth,keepaspectratio]{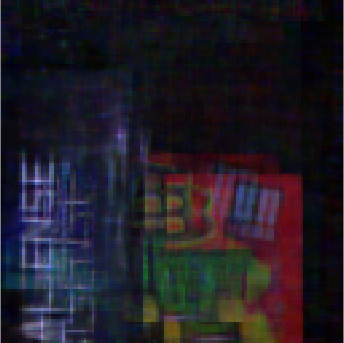} &
		\includegraphics[width=\figwidth,keepaspectratio]{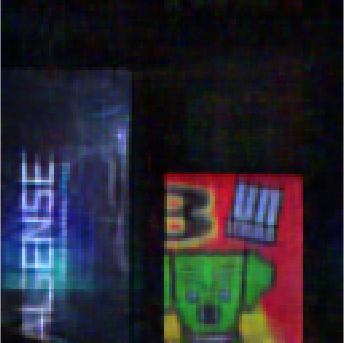} 
		\\
		\rotatebox{90}{\parbox{2.2cm}{\centering  depth}}
		& \includegraphics[width=\figwidth,keepaspectratio]{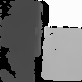} &
		\includegraphics[width=\figwidth,keepaspectratio]{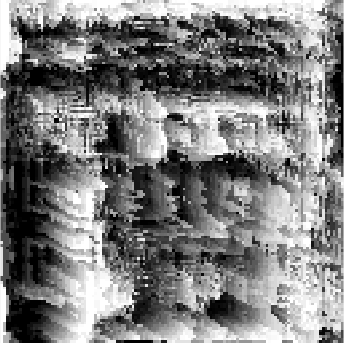} &
		\includegraphics[width=\figwidth,keepaspectratio]{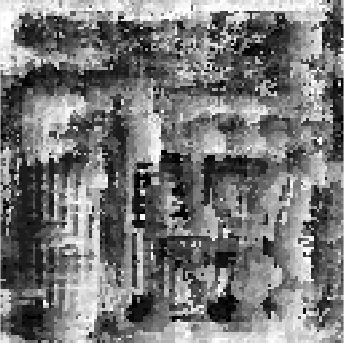}  &
		\includegraphics[width=\figwidth,keepaspectratio]{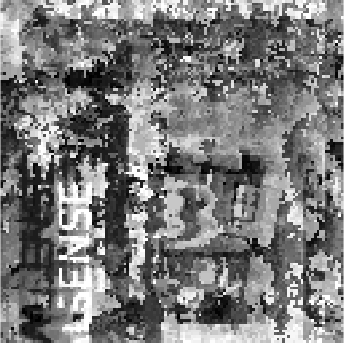}  &
		\includegraphics[width=\figwidth,keepaspectratio]{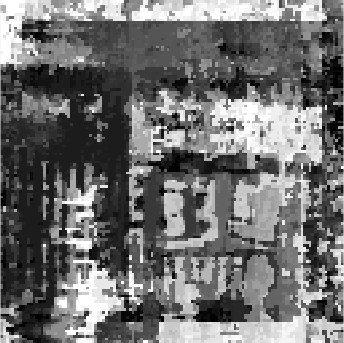} &
		\includegraphics[width=\figwidth,keepaspectratio]{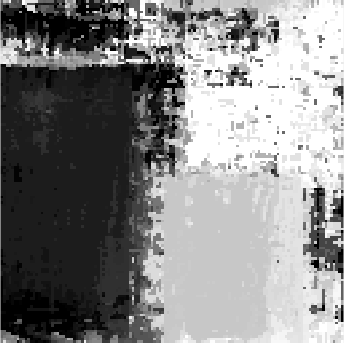} 
	\end{tabular}
    \caption{Reconstruction results  with different types and number of illumination patterns. We generate depth maps and all-in-focus images by selecting the pixel with the maximum light magnitude along each ray. The depth maps of the real scenes and the estimated depth maps are all plotted in grayscale to show range from 40cm to 60cm. We observe that 48 shifting lines provides high-quality spatial and depth resolution.}
    \label{fig:exp_illumpattern}
\end{figure*}

\subsection{Effect of Illumination Patterns}

We present experimental results of 3D reconstruction with our proposed method for real objects in Figures~\ref{fig:exp_depthplanes} and \ref{fig:exp_baselines}. 
We show the results of reconstructed depth planes, estimated all-in-focus images and depth maps using uniform, shifting lines, and shifting dots patterns. For comparison, we captured the original image and depth map for each scene using Intel RealSense D415 depth camera, where the baseline between the lens-based camera and the projector is 55mm. 

The results in Figure~\ref{fig:exp_depthplanes} compare 3D reconstruction with uniform and 48 shifting lines. In the first two rows, the scene is a slanted box, containing continuous depth varying from 40cm to 60cm. In the last two rows, the scene contains a red toy located at 40cm and a green toy lying from 50cm to 60cm. 
The results in the first three columns in Figure~\ref{fig:exp_depthplanes} represent three depth planes at 58cm, 62cm, and 66cm. The results show that the correct depth can be easily distinguished in images reconstructed with 48 shifting lines pattern, whereas depth planes reconstructed with the uniform illumination pattern show incorrect depth and intensity. The estimated all-in-focus image and depth maps for 48 shifting lines also appear significantly better than those from the uniform illumination patterns.

The results in Figure~\ref{fig:exp_illumpattern} compare different number and types of illumination patterns. We observe that the uniform illumination pattern barely recover any depth. The illumination pattern with 16 and 49 shifting dots provide better results than uniform illumination. 16 shifting lines provide slightly better results compared to shifting dots and 48 shifting lines provide significanlty better image and depth map. 

In summary, the ill-conditioned system matrices with uniform illumination pattern cause various artifacts in 3D reconstruction. Capturing measurements from coded illumination improves the conditioning of the overall system and the reconstructed images have better spatial and depth resolution. {Increasing the number of illumination patterns provides better reconstruction.} More illumination patterns would require longer acquisition time as well, which enforces a trade off between the quality of reconstruction and data acquisition time.

\renewcommand{\figwidth}{0.28\linewidth}
\begin{figure}[htb]
    \centering
        	\begin{tabular}{cccc}
		51cm &
		55cm &
		59cm
		\\
		\rotatebox{90}{\parbox{2.4cm}{\centering  5.5cm baseline}}
		\includegraphics[width=\figwidth,keepaspectratio]{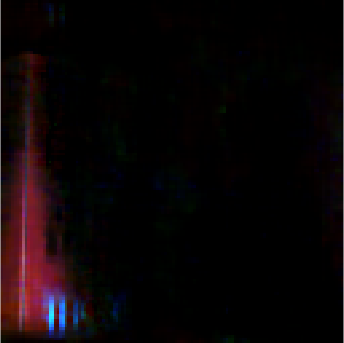} &
		\includegraphics[width=\figwidth,keepaspectratio]{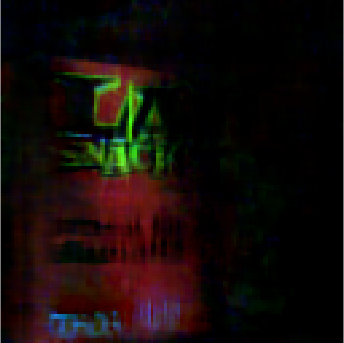} &
		\includegraphics[width=\figwidth,keepaspectratio]{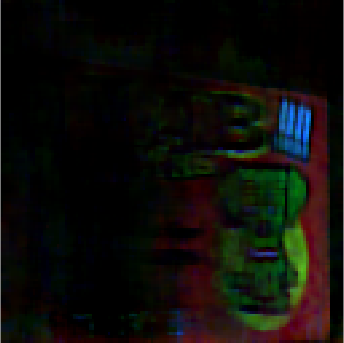} 
		\\
		\rotatebox{90}{\parbox{2.4cm}{\centering  10.5cm baseline}}
		\includegraphics[width=\figwidth,keepaspectratio]{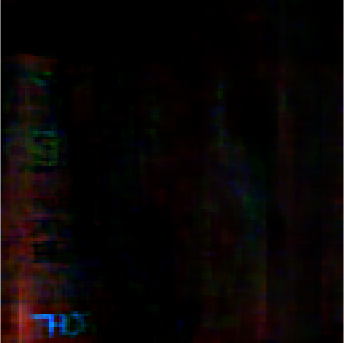}  &
		\includegraphics[width=\figwidth,keepaspectratio]{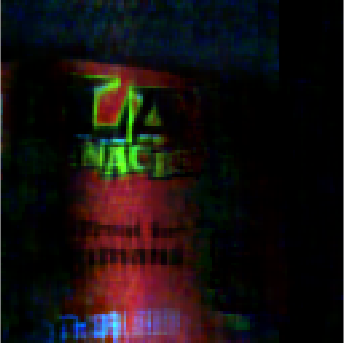} &
		\includegraphics[width=\figwidth,keepaspectratio]{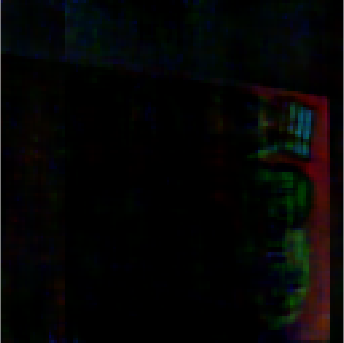} 
	\end{tabular}
    \caption{Reconstructed depth planes with 5.5cm and 10.5cm baselines within the adjustable range of hardware. The 10.5cm baseline results have better depth resolvability. The patterns are fixed with 48 shifting lines. We observe that larger baseline offers better depth resolvability.  
    }
    \label{fig:exp_baselines}
\end{figure}

\subsection{Effect of Baselines}

We show experimental results for different baselines in Figure~\ref{fig:exp_baselines}. We captured the same scene with $5.5\cm$ and $10.5\cm$ baseline and performed 3D reconstruction with the respective measurements. The results in Figure~\ref{fig:exp_baselines} show that 10.5cm baseline offers finer depth resolution (indicated as narrow depth of field) compared to the reconstruction with 5.5cm baseline. The improvement is small, and this effect was observed in the simulation results 
in Figure~\ref{fig:simu_baselines_plots} that show the depth RMSE of the system tapers off as we increase the baseline between the camera and the projector.

\section{Conclusion and Discussion}
We propose a framework for combining coded illumination with lensless imaging for 3D lensless imaging. We present simulation and real experiment results to demonstrate that our proposed method can achieve significantly improved 3D reconstruction with multiple coded illumination compared to uniform illumination. Such a mask-based lensless camera can be useful in space-limited applications such as under-the-display or large-area sensing, where installing a lens-based camera can be challenging. {Our proposed setup can also be  useful for distributed lensless sensors (in different shapes and geometries), where we may want to image over a large area, large field-of-view, but keep the devices flat, thin, and lens-free. }

\noindent \textbf{Limitaitons.} Our current setup can add extra cost and complexity to the system design because of the illumination source. The need to capture multiple shots can also increase the data acquisition time and restrict the usage for static or slow-moving objects.

\noindent \textbf{Future directions.} Extending our method to dynamic scenes is a natural direction for future work. We also need to further explore if some other illumination patters can offer better 3D reconstruction for scenes with different depth profiles. Co-design of illumination patterns, mask pattern/placement, and overall system arrangement can further improve the quality of 3D reconstruction. On the algorithmic side, the recovery algorithm can be improved by including more sophisticated priors for the 3D scenes.

\bibliographystyle{IEEEtran}
\bibliography{egbib}

\begin{IEEEbiography}[{\includegraphics[width=1in,height=1.25in,clip,keepaspectratio]{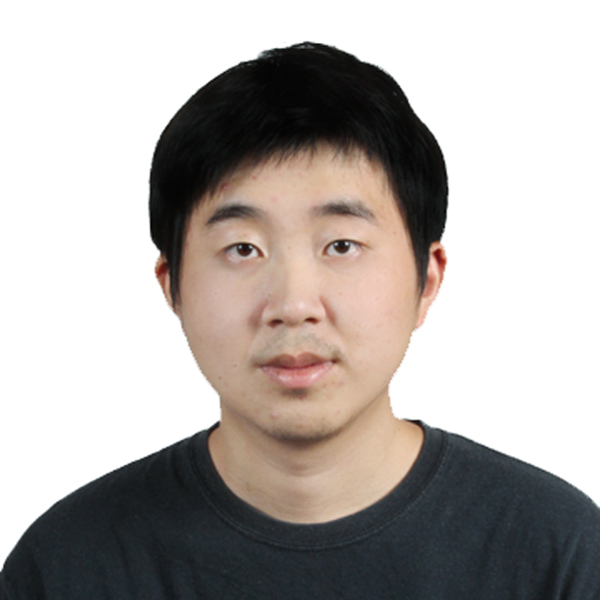}}]
	{YUCHENG ZHENG}~received the B.Sc. degree in electrical engineering from the Nanjing University of Aeronautics and Astronautics, Nanjing, China in 2017. He is currently working toward the Ph.D. degree at the University of California, Riverside, CA, USA. His current research interests include computational imaging, computer vision and signal processing.
\end{IEEEbiography}

\begin{IEEEbiography}[{\includegraphics[width=1in,height=1.25in,clip,keepaspectratio]{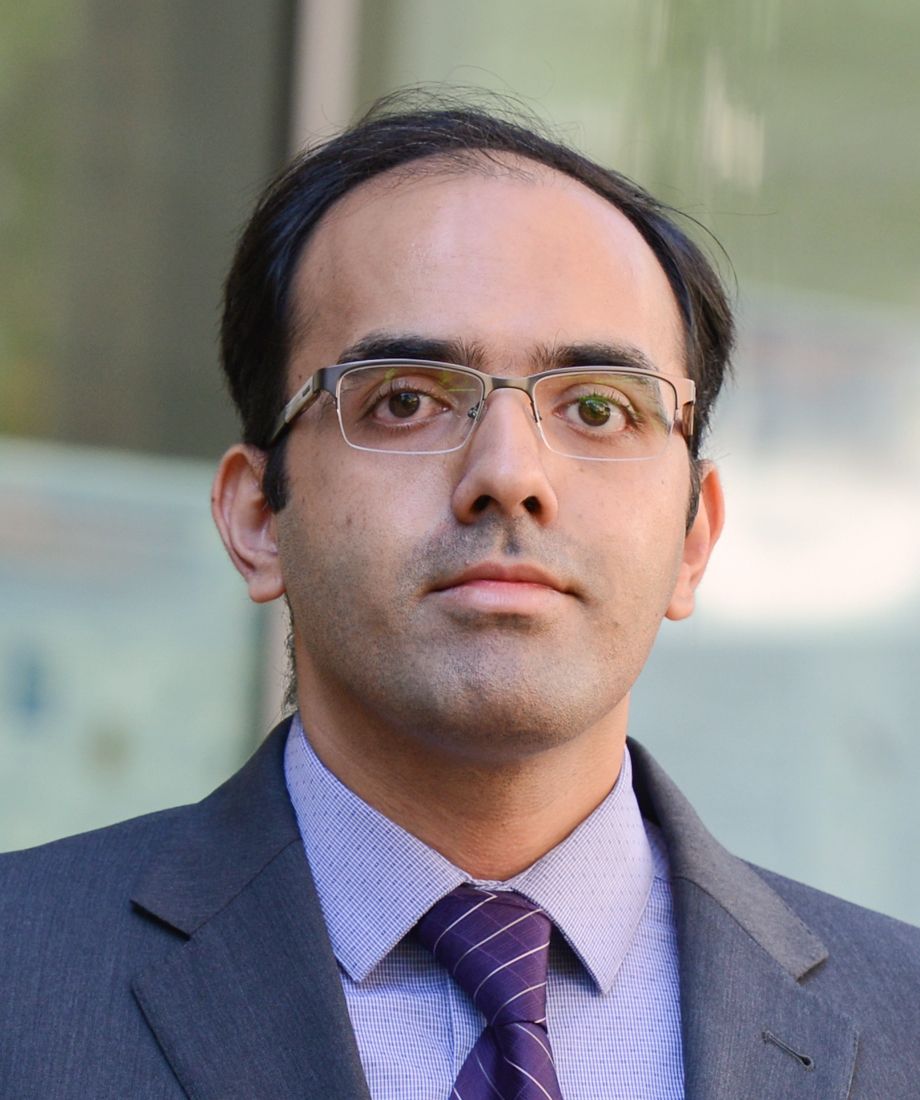}}]
	{M. SALMAN ASIF}~(Senior Member, IEEE)~received his B.Sc. degree from the University of Engineering and Technology, Lahore, Pakistan, and his M.S and Ph.D. degrees from the Georgia Institute of Technology, Atlanta, Georgia, USA. He is currently an Associate Professor at the University of California Riverside, USA. Prior to that he worked as a Postdoctoral Researcher at Rice University and a Senior Research Engineer at Samsung Research America, Dallas. He has received NSF CAREER Award, Google Faculty Award, Hershel M. Rich Outstanding Invention Award, and UC Regents Faculty Fellowship  and Development Awards. His research interests include computational imaging, signal/image processing, computer vision, and machine learning. 
\end{IEEEbiography}

\end{document}